\documentclass[sigplan,10pt]{acmart}

\usepackage{setspace}
\usepackage{amsmath, amssymb}

\usepackage[llparenthesis, rrparenthesis]{stmaryrd}

\usepackage{url}

\usepackage{semantic}

\usepackage{natbib}

\usepackage{color}
\usepackage{listings}

\usepackage{paralist}

\definecolor{mygreen}{rgb}{0,0.6,0}

\lstdefinelanguage{Lua}{
	alsoletter = -,
  keywords={function,local,end,break,if,return,nil,
  in,else,then,void,not,--,while,do,;,
  \$builtIn,
  FuncCall},
}
\newif\ifprintcomments

\newcommand{\sref}[1]{Section~\ref{sec:#1}}
\newcommand{\flabel}[1]{\label{fig:#1}}
\newcommand{\fref}[1]{Figure~\ref{fig:#1}}

\newcommand{\ie}{\emph{i.e.,} }
\newcommand{\eg}{\emph{e.g.,} }

\newcommand{\Keyword}[1]{\textsf{\textbf{#1}}}

\newcommand{\Label}[1]{{\footnotesize\textsc{#1}}}

\newcommand{\nil}{\Keyword{nil}}
\newcommand{\false}{\Keyword{false}}

\newcommand{\Variable}[1]{\textsf{\textit{#1}}}

\newcommand{\Nonterminal}[1]{\textsf{\textit{#1}}}

\newcommand{\Terminal}[1]{\textsf{\textbf{#1}}}

\newcommand{\Mterminal}[1]{$\boldsymbol{#1}$}

\newcommand{\LabelCons}[2]{\llparenthesis{\;#1\;}\rrparenthesis_{#2}}

\newcommand{\sep}{\textbar\;}

\newcommand{\dbq}[1]{``#1"}

\newcommand{\dom}[1]{\ensuremath{\mathsf{dom}(#1)}}

\begin{document}

\title[Decoding Lua: Formal Semantics\ldots]{Decoding Lua: Formal Semantics for the Developer and the Semanticist}

\author{Mallku Soldevila}
\affiliation{
  \department{FAMAF, UNC and CONICET}             
  \country{Argentina}
}
\email{mes0107@famaf.unc.edu.ar}         

\author{Beta Ziliani}
\affiliation{
  \department{FAMAF, UNC and CONICET}             
  \country{Argentina}
}
\email{bziliani@famaf.unc.edu.ar}         

\author{Bruno Silvestre}
\affiliation{
  \department{INF, UFG}            
  \country{Brazil}
}
\email{brunoos@inf.ufg.br}         

\author{Daniel Fridlender}
\affiliation{
  \department{FAMAF, UNC}             
  \country{Argentina}
}
\email{fridlend@famaf.unc.edu.ar}         

\author{Fabio Mascarenhas}
\affiliation{
  \department{DCC, UFRJ}            
  \country{Brazil}
}
\email{fabiom@dcc.ufrj.br}         

\begin{abstract}
  We provide formal semantics for a large subset of the Lua programming
  language, in its version 5.2.  We validate our model by
  mechanizing it and testing it against the test suite of the reference
  interpreter of Lua, confirming that our model accurately represents
  the language.

  In addition, we set us an ambitious goal: to target both a PL semanticist
  ---not necessarily versed in Lua---, and a Lua developer ---not necessarily
  versed in semantic frameworks.  To the former, we present the peculiarities of
  the language, and how we model them in a traditional small-step operational
  semantics, embedded within Felleisen-Hieb's \emph{reduction semantics} with evaluation
  contexts. The mechanization is, naturally, performed in PLT Redex, the \emph{de
    facto} tool for mechanizing reduction semantics.

  To the reader unfamiliar with such concepts, we provide, to our best possible
  within the space limitations, a gentle introduction of the model. It is our
  hope that developers of the different Lua implementations and dialects
  understand the model and consider it both for testing their work and for
  experimenting with new language features.
\end{abstract}

\maketitle
\renewcommand{\shortauthors}{Soldevila et al.}

\section{Introduction}

Lua is a lightweight imperative scripting language,
featuring dynamic typing, automatic memory management, data description 
facilities, and metaprogramming mechanisms to adapt the language to specific 
domains \cite{spe}. 
The typical use case of a Lua application is as an extension library embedded 
in a \textit{host} application, commonly written in C or C++. In that setting, 
Lua offers the possibility to add scripting facilities to the host
application, combining the flexibility and rapid prototyping of 
a dynamic language within the static guarantees and
optimizations of stricter programming languages.

Lua is extensively used in many diverse
applications, ranging from game development, most notably by ``AAA"
games~\cite{ieru01} but also in mobile games and game frameworks,
plugin development (for example, in the photo editing software Adobe
Photoshop
Lightroom\footnote{\url{http://www.adobe.com/devnet/photoshoplightroom.html}},
and the type-setting system
LuaTex\footnote{\url{http://www.luatex.org/languages.html}}), 
web application firewalls\footnote{\url{https://blog.cloudflare.com/cloudflares-new-waf-compiling-to-lua}}, and embedded systems\footnote{\url{https://www.lua.org/uses.html}}.

Lua is informally specified by both its reference manual and its reference
interpreter, developed and maintained by the core Lua authors. Thanks to Lua's success, several
alternative 
implementations\footnote{\url{http://lua-users.org/wiki/LuaImplementations}},
as well as code linters and static 
analyzers\footnote{\url{http://lua-users.org/wiki/ProgramAnalysis}} can be
found in the wild. However, the informal nature of the specification
means that developers of those tools must resort to their intuition,
formed by study of the reference manual, inspection of the source code of the 
interpreter, and experimentation.

In this work, we present a comprehensive formalization of (most of)
Lua 5.2, which we argue will facilitate the development and testing
of these alternative implementations and analysis tools, as well
as the prototyping of new features and extensions to the Lua language.

The formalism we use to express the semantics of Lua is mainly Felleisen-Hieb's
reduction semantics with evaluation contexts~\cite{syntheo}, better known as
just \emph{reduction semantics}. This formalism, together with
concepts from standard small-step operational semantics, 
has already been proven useful for tackling the formalization of real
programming languages \cite{guh, python, get, semcontext}.

While providing the semantics on paper is an important contribution,
it does not suffice to ensure that our characterization of the language is 
correct. For this reason
we mechanize the semantics in PLT Redex \cite{plt}, following the success of
previously mechanized reduction semantics for other scripting languages such as
JavaScript~\cite{guh} and Python~\cite{python}.

We tested the mechanization of our formal semantics of Lua against the test suite of the reference interpreter, successfully passing every test within the 
scope of the formalization.
We take this as strong evidence to support the claim that our
semantics is a sound representation of the selected subset of the language's
features, including:
\begin{compactitem}
\item Every type of Lua value, except {\em coroutines} and {\em userdata} (see below);
\item Metatables;
\item Identity of closures;
\item Dynamic execution of source code;
\item Error handling (throwing error objects and catching them with {\em protected mode}).
\item A large collection of the services of the standard library.
\end{compactitem}
We purposely left out the following features for future work:
\begin{compactitem}
\item Coroutines, in essence single-shot delimited continuations;
\item Userdata, opaque handles to data from the host application and native
libraries;
\item Garbage collection;
\item The \Keyword{goto} statement;
\item Services from the standard library that interface with the operating
system, such as file manipulation, or have large complex C implementations,
such as string pattern matching.
\end{compactitem}

Our work is heavily motivated by previous work 
formalizing and mechanizing other scripting languages~\cite{guh,python}. 
However, we take a different path, motivated by the specificities of Lua.
The most important difference is that we do not attempt to reduce
the language to a minimal core, and then specify a {\em desugaring}
or elaboration step from source programs to this minimal core. 
We keep most of the original surface syntax of Lua, with minimal
changes. Our goal is to make the semantics and its traces understandable by
programmers that are already familiar with the Lua language but
are not used to the concepts of reduction semantics that form
the core of our formalization.

Another important difference is that we keep a strict separation between the
semantic model (what forms the core of what is presented in this work) and its
mechanization. While this is a rather technical point, it has an impact in the
model, which we will timely discuss.

The rest of the paper is organized as follows: \sref{langintro}
presents a very brief description of Lua, with emphasis on some
of the features that we formalize in later sections;
\sref{formalism_used} presents the basic concepts that
our formalization uses, via a formalization of a very small subset
of Lua; \sref{syntax_semantics} expands the formalization
of the previous section to show the most important parts
of our complete semantics; \sref{mechanization} briefly
discusses the mechanization and its tests; \sref{relatedwork}
discusses related work; finally, \sref{conclusion}
summarizes our contributions and discusses future avenues
of research.

\section{Lua, an extensible scripting language}\label{sec:langintro}

We organize our presentation of Lua around the examples of memoization
and object-oriented programming, shown in figures
\ref{lst:example:memoization} and \ref{lst:example:oo},
respectively. They serve to introduce several characteristics of Lua:
its syntax, the versatility of its single data structure (\textit{tables}),
its metaprogramming mechanisms and some aspects of its scoping rules.

\subsection{Memoization}

\begin{figure}
\begin{lstlisting}
local function memoize(fn)
  local t = {}
  return function(x)
    local y = t[x]
    if y == nil then y = fn(x); t[x] = y end
    return y
  end
end

local memsum = memoize(function(x)
  local a = 1
  for i = 1,x do a = a + i end
  return a
end)
\end{lstlisting}
\caption{Memoization in Lua.}
\label{lst:example:memoization}
\end{figure}

The code\footnote{Taken from
  \url{http://lua-users.org/wiki/FuncTables}\,.} shown in
Figure~\ref{lst:example:memoization} implements a memoization
function, \textsf{memoize}, which takes a function \textsf{fn} as
argument and returns its \emph{memoized} version. The values of
\textsf{fn} already computed will be stored in a \textit{table} (\textsf{t} in line 2). At their core, tables are
associative arrays that can be indexed with any Lua
value except \nil. We will show later in this section
that tables also come with syntax sugar and metaprogramming
facilities that can greatly extend their functionality beyond simple
associative arrays.

Line 3 is where the memoized version of \textsf{fn} is
returned through an anonymous function. 
This function takes \textsf{x} as argument and, before
computing \textsf{fn(x)}, performs a look-up in the table for value
\textsf{x} (line 4). If the result of the look-up is \nil{} it means no result was found, so it proceeds to compute \textsf{fn(x)} 
and store it in the table (line 5).
The resulting value, either computed or retrieved from the table, is
returned in line 6. The function \textsf{memoize} is used in lines 10--14 to improve the performance of a function that performs a sum from 1 to \textsf{x}. 

All procedures and functions in Lua, anonymous or named, are first-class values,
and form lexically-scoped closures. The anonymous function that \textsf{memoize}
returns will effectively capture into its scope the table \textsf{t}, as
expected.

Note that the definitions of \textsf{memoize}, \textsf{t}, and \textsf{memsum} are prefixed by the keyword
\textsf{\textbf{local}}. Without it, all of these
declarations are simple assignments, and do not introduce new
names in the current scope. In an assignment, if there is no
variable in scope with that name, then the variable is global:
the assignment will actually store its rvalue in a table
called the \textit{environment}, with a string containing the
variable's name as the key. Using a variable that is not in scope
also looks up the variable in the environment. 

The environment is available to the programmer through a \emph{variable}
\textsf{\_ENV}, which is always in scope. This means that any occurrence of a
variable $x$ that is \emph{not} in scope is just syntax sugar for
\textsf{\_ENV[\dbq{$x$}]}.  Since it is a variable, the programmer can change the
environment at will by simply assigning another table to \textsf{\_ENV}.

\subsection{Simple OOP in Lua}

\begin{figure}
\begin{lstlisting}
local MyClass = {}
MyClass.__index = MyClass

function MyClass.new(init)
  local self = setmetatable({}, MyClass)
  self.value = init
  return self
end

function MyClass:set_value(newval)
  self.value = newval
end

function MyClass:get_value()
  return self.value
end

local mc = MyClass.new(5)
print(mc:get_value()) -->> 5
mc:set_value(6)
print(mc:get_value()) -->> 6
\end{lstlisting}
\caption{OOP based on Lua's metatable mechanism.}
\label{lst:example:oo}
\end{figure}

Another interesting example\footnote{Taken from
  \url{http://lua-users.org/wiki/ObjectOrientationTutorial}\,.} is
listed in Figure~\ref{lst:example:oo}. It presents the implementation
of some basics concepts of object-oriented programming, namely classes
and objects, by combining tables, first-class functions, and the
\textit{metatable} mechanism. It also presents some syntax sugar provided by Lua to better support OOP.

In Lua, a class is essentially implemented as a dictionary (\eg table), in which
the method names form the keys of the dictionary, and the method implementations
are the associated values.  Objects are also model with tables, containing the
fields and their values.

In the example, we have a class \textsf{MyClass} with
its corresponding constructor (line 4) and only one field
\textsf{value} with its setter (line 10) and getter (line 14).
The function declarations in these lines are actually syntax
sugar for assignments, where the left-hand sides
are, respectively, \textsf{MyClass[\dbq{new}]}, \textsf{MyClass[\dbq{set\_value}]}, and \mbox{\textsf{MyClass[\dbq{get\_value}]}}. For the two methods on line 10 and
line 14 the use of \textsf{:} instead of \textsf{.} also includes
an extra first parameter for these functions, named \textsf{self}.

In the last lines of Figure~\ref{lst:example:oo} we show how to create
an instance of \textsf{MyClass} (line 18), and how to invoke the
methods. In line 20 we can observe the invocation of
\textsf{set\_value} with yet another syntax sugar:
\textsf{mc:set\_value(6)} is equivalent to
\textsf{mc[\dbq{set\_value}](mc, 6)}.

If classes contain methods, and objects contain fields, how is
\textsf{mc[\dbq{set\_value}]} looking up the \textsf{set\_value} method? The
answer is the \textit{metatable} mechanism, used in lines 2 and 5. In line 5,
the call to \textsf{setmetatable} assigns \textsf{MyClass} as the metatable of
the empty table \textsf{\{\}} passed as argument, and then returns this empty
table.

A metatable can modify the behavior of a table with regards
to most of Lua's operations. For this example, the behavior
we are modifying is look-up of non-existing keys. Each
behavior that can be modified has an associated handler.
For look-up of non-existing keys the handler is called
\textsf{\_\_index} (line 2). A handler is usually a function, but
in the case of \textsf{\_\_index} it can be another table, in this case \textsf{MyClass}.
A non-existing key then will be looked up in this table, and
this is how \textsf{mc[\dbq{set\_value}]} results in 
the method \textsf{set\_value} from \textsf{MyClass}.

Lua also specifies handlers for setting a non-existing key,
for calling a value as if it were a function, for
most of the binary and unary operators, for setting finalizers, 
and even for some functions in the standard library. Lua programmers
typically use metatables for object-oriented programming
(including more elaborated object models than class-based
single inheritance), for operator overloading, and for
proxies.

\vspace{-.5em}
\section{Basics of the formalization}
\label{sec:formalism_used}

In this section, we gently introduce the semantic framework used throughout the paper by providing semantics to a small subset of Lua. Essentially, we mix classical ideas from operational semantics based on
abstract machines ---the notion of ``operational'', how a
program performs its computations, and the representation of a \emph{store} or memory--- together
with reduction semantics with evaluation contexts~\cite{plt}, formalism from which we take the tools for modeling the concepts of environment and continuations, and to obtain a modular description of the semantics from simple computations to the execution of complete programs.

An interesting aspect of Felleisen-Hieb's reduction semantics is the
possibility of defining the semantics of a language by decomposing it into
fragments, describing the fragment's semantics in isolation with a separate
relation. For our small subset of the Lua language, we describe three fragments: \emph{pure} statements, pure expressions (following Lua's distinction
of statements and expressions), and \emph{stateful} (\ie memory changing) statements. Then, we compose
the three using a fourth relation,
therefore finally providing the semantics for entire programs.

\begin{figure}
  \begin{tabbing}
        \Nonterminal{s} ::= \= \Keyword{if} \Nonterminal{e} \Keyword{then} \Nonterminal{s}
                               \Keyword{else} \Nonterminal{s} \Keyword{end}
                          \sep \Terminal{;}\\[.3em]
       \Nonterminal{v} ::= \= \Keyword{nil} \sep \Nonterminal{bool\_literal}\\[.3em]
       \Nonterminal{e} ::= \= \Nonterminal{v} \sep
                              \Nonterminal{e} \Nonterminal{binop} \Nonterminal{e}
                              \sep \Nonterminal{unop} \Nonterminal{e}\\[.3em]
       \Nonterminal{binop} ::= \= \Keyword{and} \sep \Keyword{or}\\[.3em]
       \Nonterminal{unop} ::= \= \Keyword{not}
        \end{tabbing}
\caption{Syntax of simple statements and expressions}
\flabel{model_language_syntax}
\end{figure}

\begin{figure}
           $$\setnamespace{2em}
				\inference{\Variable{v} \notin \{ \Keyword{nil}, \Keyword{false} \}}
			    {\mathsf{\Keyword{if}\;\Variable{v}\;\Keyword{then}\;
			                \Variable{s}_1\;
			             \Keyword{else}\;
			                \Variable{s}_2\;
			             \Keyword{end}\;
			        \rightarrow ^{\textsf{s}}\;
			            \Variable{s}_1}}
			            []$$
			
			$$\setnamespace{2em}
			    \inference{\Variable{v} = \Keyword{nil} 
			                        \vee 
			                \Variable{v} = \Keyword{false}}
			    {\mathsf{\Keyword{if}\;\Variable{v}\;\Keyword{then}\;
			                \Variable{s}_1\;
			             \Keyword{else}\;
			                \Variable{s}_2\;
			             \Keyword{end}\;
			         \rightarrow ^{\textsf{s}}\;
			             \Variable{s}_2}}
			             []$$
\caption{Semantics of the conditional statement.}
\flabel{model_conditional}
\end{figure}

\begin{figure}
\begin{tabular}{cc}
$\Keyword{not}\;\Variable{v}
\rightarrow^e\;
\boldsymbol{\delta}(\Keyword{not}, \Variable{v})$ &

			$$\setnamespace{0.5em}
			\inference
			{\Nonterminal{op} \in \{ \Keyword{and}, \Keyword{or} \}}
			{\Variable{v}\;\Variable{op}\;\Variable{e}\;
			    \rightarrow^e\;
			 \boldsymbol{\delta}(\Variable{op}, \Variable{v}, \Variable{e})}[]$$
\end{tabular}
\caption{Semantics of expressions.}
\flabel{model_simple_expressions}
\end{figure}

\begin{figure}
	\begin{tabbing}
		$\mathsf{\boldsymbol{\delta}\textsf{(\Keyword{and},\;\Variable{v},\;\Variable{e})}}$ \= $= 
		\left\{ \begin{array}{lcl}
		\textsf{v} & \textsf{\textbf{if}}\;\Variable{v}\;=\;\Keyword{false} \vee \Variable{v}\;=\;\nil
		\\
		\Variable{e} & \textsf{\textbf{otherwise}}
		\end{array}
		\right.\;\;\;\;\;\;\;\;\;\;\;\;\;\;\;\;\;\;\;\;\;\;\;\;\;\;\;\;\;\;\;\;\;\;
		$\\[.5em]
				
		$\mathsf{\boldsymbol{\delta}\textsf{(\Keyword{or},\;\Variable{v},\;\Variable{e})}}$ \> $= 
		\left\{ \begin{array}{lcl}
		\textsf{v} & \textsf{\textbf{if}}\;\Variable{v}\;\neq\;\Keyword{false} \wedge 
							\Variable{v}\;\neq\;\Keyword{nil}
		\\
		\Variable{e} & \textsf{\textbf{otherwise}}
		\end{array}
		\right.\;\;\;\;\;\;\;\;\;\;\;\;\;\;\;\;\;\;\;\;\;\;\;\;\;\;\;\;\;\;\;\;\;\;
		$\\[.5em]
		
		$\mathsf{\boldsymbol{\delta}\textsf{(\Keyword{not},\;\Variable{v})}}$ \> $= 
		\left\{ \begin{array}{lcl}
		\Keyword{true} & \textsf{\textbf{if}}\;\Variable{v}\;=\;\Keyword{false} \vee 
							\Variable{v}\;=\;\nil
		\\
		\Keyword{false} & \textsf{\textbf{otherwise}}
		\end{array}
		\right.\;\;\;\;\;\;\;\;\;\;\;\;\;\;\;\;\;\;\;\;\;\;\;\;\;\;\;\;\;\;\;\;\;\;
		$
	\end{tabbing}
\caption{$\boldsymbol{\delta}$ function: boolean operators.}
\flabel{delta_function_bool_op}
\end{figure}

We show the grammar for \emph{stateless} programs in
\fref{model_language_syntax}. The statements are conditional branching and skip
(denoted with \textsf{;}\,). The expressions are \nil\; (the absence of a useful
value), boolean constants, and logical operators. Of course, we are not able to
write any useful program. In the coming sections we will grow our language until
we reach Lua.

\fref{model_conditional} introduces the typical operational semantics for the
conditional statement, modeled with the $\rightarrow ^{\textsf{s}}$ relation 
between stateless statements. The first rule states that, in
a boolean context (the conditional of the \Keyword{if}), any value different
from \nil\; and \Keyword{false} is considered \Keyword{true}, and therefore the
\Keyword{then} branch is considered. Note that we write above the line the
conditions in which the rule applies. When no condition is required, the line will be omitted. The second rule states that, for \Keyword{false} or \Keyword{nil}, the \Keyword{else} branch is considered.

\fref{model_simple_expressions} gives the semantics of expressions using a
separate $\rightarrow^e$ relation. We use an \textit{interpretation} function
$\delta$, as seen in the literature, which provides meaning to operators using
\emph{denotational} semantics. In contrast to the relations over terms presented so far, denotational semantics are not tight to single computation steps. \fref{delta_function_bool_op} shows (a simplified version of\footnote{The actual
  equations use Lua's \emph{parenthesized expressions}, introduced in
  \ref{sec:stat_expr_no_store}.}) the $\delta$-equations for boolean operators.

\begin{figure}
	$$\setnamespace{0.5em}
			\inference*{\mathsf{\sigma\textsf{'} =
								(\Variable{r},\Variable{v}),\sigma}}
						{\mathsf{\begin{aligned}\sigma\;\textbf{:}\;
							\Keyword{local}\;
								\Variable{x}\;
											& \textbf{=}\;
								\Variable{v}\;
							\Keyword{in}\;
								\Variable{s}\;
							\Keyword{end}\;
								\rightarrow^{{\sf s\_\sigma}}\;
							\sigma\textsf{'}\;\textbf{:}
								\;\Variable{s}[\Variable{x} \backslash \Variable{r}]\end{aligned}}}
			[]$$\\
\begin{tabular}{cc}			
    $$\setnamespace{0.5em}
			\inference*{\mathsf{\sigma\textsf{'} = \sigma [r := \Variable{v}]}}
						{\mathsf{\sigma\;\textbf{:} \;\Variable{r}\;\Terminal{=}\;\Variable{v}\;
							\rightarrow^{{\sf s\_\sigma}}\;
								\sigma\textsf{'}\;\textbf{:}\;\Terminal{;}}}
		[]$$\;\; &\;\;
	$\sigma\;\textbf{:}\;r \rightarrow^{{\sf e\_\sigma}}\; \sigma\;\textbf{:}\;\sigma(r)$
	\end{tabular}\\[.3em]
\caption{Semantics of variables and references.}
\flabel{model_semantics_statements_sigma}
\end{figure}

We proceed now to extend the language with imperative features: (local) variables. Statements are enlarged with variable definition and assignment:
\[        \Nonterminal{s} ::=  ... ~|~ 
                                \Keyword{local} 
			                         \;\Nonterminal{x} \;\Terminal{=} \;
			                     \Nonterminal{e}  
			                     \;\Keyword{in}\;
			                     \Nonterminal{s} \;
			                     \Keyword{end} ~|~
                              \Nonterminal{x} \;\Terminal{=}\; \Nonterminal{e}
\]
  In order to describe its operational semantics, we must introduce a model of
  the memory store. We model it as a partial function from a set of references
  to values, denoted as $\sigma$. We refer to $\sigma$ as the ``\textit{values'
    store}'', or simply \emph{store}.

As for references, we will not force any specific representation, just ask them
to satisfy some simple properties to ensure the relation modeling the semantics
of variables stays decidable. More
specifically, we ask the domain of $\sigma$ (referred as to
\dom{\sigma}) to be a finite set, with elements that must be
syntactically represented, but different from any other syntactic object
in the language. We further assume it is 
always possible to obtain a fresh reference from the store.

We extend the grammar of expressions with references:
\[
 \Nonterminal{e} ::= \;\;... ~|~ \Nonterminal{r}
\]
References, in contrast to all the language constructs we mentioned so far, do not belong to the Lua source language, \ie they cannot be written down by a developer.
They are
\emph{run-time constructs}: syntactic extensions made to the language for the sole purpose of
obtaining a small-step semantics of the language.
We will see other examples
of such constructs in the coming sections.

\fref{model_semantics_statements_sigma} describes the semantics for the
definition and assignment of local variables. We use a new $\rightarrow^{{\sf s\_\sigma}}$ relation, which maps a pair of a store $\sigma$ and a statement $\mathsf{s}$ with another pair of 
a new store $\sigma'$ and the resulting statement $\mathsf{s}'$.

As shown in the rule for the introduction of local variables, when the right side of the definition is a value $\Variable{v}$
we put it in the store with a fresh reference $\Variable{r}$. Then, we replace each occurrence of variable $\Variable{x}$ in the scope of
the {\bf local} statement by the new reference $\Variable{r}$.

An important property of this semantics is that variables are never
free, as substitution will always replace them by references right before they would become free.
This will have an impact on closure creation (see 
\ref{sec:stat_expr_store}).

Returning to \fref{model_semantics_statements_sigma}, the second rule shows
variable assignment, with $\sigma [\Variable{r} := \Variable{v}]$ denoting a store
$\sigma'$ such that $\dom{\sigma'} = \dom{\sigma}$, where
$\sigma'(\Variable{r}) = \Variable{v}$ and
$\forall \Variable{r'} \in \dom{\sigma'}, \Variable{r'} 
\neq \Variable{r} \Rightarrow \sigma'(\Variable{r'}) = \sigma(\Variable{r'})$.  
Note that the assignment reduces to an empty statement \Terminal{;},
indicating that there is nothing else to do for this particular
statement. The third and final rule shows that references appearing in expressions are always implicitly dereferenced.

\begin{figure}
\begin{tabbing}
\Nonterminal{E} ::= \= \;\;[\;]
			     \sep \Keyword{if} \Nonterminal{E} \Keyword{then} 
			                     \Nonterminal{s} 
			                  \Keyword{else} 
			                     \Nonterminal{s} 
			                  \Keyword{end}\\
			    \> \sep \Keyword{local} \Nonterminal{x} \Terminal{=} E 
							   \Keyword{in} \Nonterminal{s} 
							   \Keyword{end}
				\sep \\
			    \> \sep \Nonterminal{x} \Terminal{=} \Nonterminal{E} 
			     \sep \Nonterminal{E} \Nonterminal{binop} \Nonterminal{e}
			     \sep \Nonterminal{v} \Nonterminal{binop} \Nonterminal{E}
			     \sep \Nonterminal{unop} \Nonterminal{E}
		\end{tabbing}
\caption{Evaluation contexts.}
\flabel{model_ev_contexts}
\end{figure}

\begin{figure}
    \begin{tabular}{cc}
		$$\setnamespace{2em}
			\inference*{\mathsf{\Variable{e}\;\rightarrow^e\;\Variable{e}'}}
			{\mathsf{\sigma\; \textbf{:} 
						\;\Variable{E} [\![ \Variable{e} ]\!]\;
							\mapsto\;
					\sigma\; \textbf{:} \; \Variable{E} [\![ \Variable{e}' ]\!]}}
		[] $$ &

		$$\setnamespace{2em}
			\inference{\mathsf{\Variable{s}\;
							\rightarrow^{s}\;
							\Variable{s}'}}
			{\mathsf{\sigma\; \textbf{:}
						\;\Variable{E} [\![ \Variable{s} ]\!]\;
							\mapsto\;
					\sigma\; \textbf{:} \; \Variable{E} [\![ \Variable{s}' ]\!]}}
		[] $$
\end{tabular}
\begin{tabular}{cc}
\\		$$\setnamespace{2em}
			\inference{\mathsf{\sigma\; \textbf{:}\; \Variable{s}\;
							\rightarrow^{s\_\sigma}\;
							\sigma'\; \textbf{:}\; \Variable{s}'}}
			{\mathsf{\sigma\; \textbf{:}
						\;\Variable{E} [\![ \Variable{s} ]\!]\;
							\mapsto\;
					\sigma'\; \textbf{:}\; \Variable{E} [\![ \Variable{s}' ]\!]}}
		[]$$ &
		$$\setnamespace{2em}
		\inference{\mathsf{\sigma\; \textbf{:}\; \Variable{e}\;
				\rightarrow^{e\_\sigma}\;
				\sigma'\; \textbf{:}\; \Variable{e}'}}
		{\mathsf{\sigma\; \textbf{:}
				\;\Variable{E} [\![ \Variable{e} ]\!]\;
				\mapsto\;
				\sigma'\; \textbf{:}\; \Variable{E} [\![ \Variable{e}' ]\!]}}
		[] $$		
\end{tabular}\\[.4em]
\caption{Semantics of programs.}
\flabel{model_standard_red}
\end{figure}

We have already defined three different relations, each of them
computing a bit of a program: $\rightarrow ^{\textsf{e}}$ computes an
expression, $\rightarrow ^{\textsf{s}}$ an stateless statement, and
$\rightarrow^{{\sf s\_\sigma}}$ an stateful statement. Now we are ready to
combine the three to perform the execution of a full program. To that effect we define
the $\mapsto$ relation. This relation will
say exactly when each of the previously defined relations will trigger, at the
same time defining the order in which statement or expression must be executed
next.

Here is where \emph{evaluation contexts} play a central role. They describe the
syntax of the language with the addition of
a new construction: a \textit{hole}, usually denoted as \textsf{[ ]}. Evaluation
contexts will play different roles in later sections, but for the moment the (only)
hole in a program will be filled in with the next statement or expression to be
executed.

\fref{model_ev_contexts} defines the evaluation context $\Nonterminal{E}$ for the small subset of Lua we
described so far. We can see from the definition the order we expect evaluation to take place: in an \Keyword{if}, the guard must be evaluated first. In the definition of variables we evaluate the
rvalue for the definition first. In a binary operation, we evaluate the left operand first\footnote{Our definition enforces left-to-right
	evaluation of expressions. Even if this is left unspecified in
	Lua's reference manual, the two most popular implementations of Lua, the reference
	interpreter and LuaJIT (\url{luajit.org}), both evaluate expressions
	left-to-right.}.

\fref{model_standard_red} defines the $\mapsto$ relation. Both
$\Variable{E} [\![ \Variable{e} ]\!]$ and $\Variable{E} [\![ \Variable{s} ]\!]$ denote an evaluation context where the hole is filled with
the respective expression or statement, if this yields a 
syntactically valid term of the language. If the evaluation context is well-defined, together with
the relations that describes computation steps, there is a
unique decomposition of a valid term into an evaluation context and
a subterm, and this subterm will match one and only one of the semantic rules.
The subterm that is filling the hole gives the current focus of
the computation.

With all of the main ingredients in place, we are now ready to provide semantics to Lua.

\section{A formal description of Lua}
\label{sec:syntax_semantics}

In this section, we describe the highlights of our formalization of the
semantics of Lua, the main contribution of this work. 
\sref{stat_expr_no_store} covers the stateless subset of the language, 
\sref{stat_expr_store} covers the imperative subset, \sref{built_in_sec}
describes the concepts added to support standard library services, \sref{metatables} covers
the semantics of metatables, and \sref{semantics_programs} wraps up
with the semantics of complete
programs and error handling.

\subsection{Stateless Lua}
\label{sec:stat_expr_no_store}

\begin{figure}
  \begin{tabbing}
        \Nonterminal{s} ::= \= ... \sep \Keyword{while} \Nonterminal{e} \Keyword{do}
						                    \Nonterminal{s}
						                 \Keyword{end}
						  \sep \Keyword{break}
                          \sep \Nonterminal{s} \Nonterminal{s}\\

       \Nonterminal{v} ::= \= ... \sep \Nonterminal{number\_literal}
		            \sep \Nonterminal{string\_literal}\\

       \Nonterminal{binop} ::= \= ... \sep \Nonterminal{strictbinop}\\

       \Nonterminal{strictbinop} ::= \={\small \Terminal{+} \sep \Terminal{-}
		                           \sep \Terminal{*} \sep \Terminal{/}
		                           \sep \Mterminal{\hat{}}
						 	\sep \Terminal{\%} \sep \Terminal{..} \sep \Mterminal{<}
							\sep \Mterminal{\leq} \sep \Mterminal{>} 
						  \sep \Mterminal{\geq} \sep \Terminal{==}}\\

        \Nonterminal{unop} ::=\;\;... \sep \Terminal{-} \sep \Terminal{\#}
        \end{tabbing}
\caption{Syntax of the remaining stateless subset.}
\flabel{syntax_simple}
\end{figure}

\begin{figure}
  \begin{tabbing}
    \Nonterminal{s} ::= \= ... \sep \Keyword{\$iter} \Nonterminal{e} \Keyword{do} \Nonterminal{s} \Keyword{end}
      \sep $\LabelCons{\Nonterminal{s}} {\Nonterminal{label}}$\\
    \Nonterminal{label} ::= \=\Label{Break}
   \end{tabbing}
\caption{Run-time statements for \Keyword{while} and 
 \Keyword{break}.}
\flabel{while_runtime}
\end{figure}

\begin{figure}
\begin{align*}
  \Keyword{while}\;\Variable{e}\;\Keyword{do}\;\Variable{s}\;\Keyword{end}
  &\rightarrow^{\textsf{s}}
  \LabelCons{\Keyword{\$iter}\;\Variable{e}\;\Keyword{do}\;\Variable{s}\;\Keyword{end}}{\Label{Break}} \\
  \Keyword{\$iter}\;\Variable{e}\;\Keyword{do}\;\Variable{s}\;\Keyword{end}
  &\rightarrow^{\textsf{s}}
  \Keyword{if}\;\Variable{e}\;\Keyword{then}\;\Variable{s}\;\Keyword{\$iter}\;\Variable{e}\;\Keyword{do}\;\Variable{s}\;\Keyword{end}\\&\;\;\;\;\;\;\;
  \Keyword{else}\;\Terminal{;}\;\Keyword{end} \\
  \Terminal{;}\;\Variable{s} &\rightarrow^{\textsf{s}} \Variable{s}\\
  \LabelCons{\Variable{E}_\textsf{lf} [\![\;\Keyword{break} \;]\!]}{\Label{Break}}&\rightarrow ^{\textsf{s}}\Terminal{;}\\
  \LabelCons{\Terminal{;}}{\Label{Break}}&\rightarrow^{\textsf{s}}\Terminal{;}
\end{align*}		
\caption{Semantics of stateless statements.}
\flabel{semantics_simple_statements}
\end{figure}

\begin{figure}
  \begin{tabbing}
    \Nonterminal{e} ::= \= ... \sep $\LabelCons{\Nonterminal{e}}
    {\Nonterminal{label}}$  \\
    \Nonterminal{label} ::= \= ... \sep \Label{ArithWO}
    \sep \Label{ConcatWO} \sep \Label{OrdWO} \sep \ldots
  \end{tabbing}
\caption{Run-time expressions for errors.}
\flabel{runtime_constructions_exps_no_store}
\end{figure}

\begin{figure}
			$$\setnamespace{0.5em}
			\inference
			{\Variable{op} \in \{ \textbf{+}, \textbf{-}, \textbf{*}, \textbf{/},
			                       \;\boldsymbol{\hat{\null}}\;, \textbf{\%} ,
			                       \boldsymbol{<} , \boldsymbol{\leq} \}
			 \;\;\;\;\Variable{v}_1, \Variable{v}_2\;\in\;\Nonterminal{number}}
			{\Variable{v}_1\;\Variable{op}\;\Variable{v}_2\;
			    \rightarrow^e\;
			 \boldsymbol{\delta}(\Variable{op}, \Variable{v}_1, \Variable{v}_2)
			}
			[]$$

			$$
			\setnamespace{0.5em}
				\inference
				{\Variable{op} \in \{ \textbf{+}, \textbf{-}, \textbf{*}, \textbf{/}, \boldsymbol{\hat{\null}}\;,
				                                \textbf{\%}\}\\
				 \Variable{v}_1\;\notin\;\Nonterminal{number}\;
				                 \vee\;
				 \Variable{v}_2\;\notin\;\Nonterminal{number}\\
			     \nu_1 = \boldsymbol{\delta}(\Keyword{tonumber}, \Variable{v}_1, 10) \; 
			     \in\;\Nonterminal{number}\\
			     \nu_2 =  \boldsymbol{\delta}(\Keyword{tonumber}, \Variable{v}_2, 10)\;
				                 \in\;\Nonterminal{number}
				}
				{\Variable{v}_1\;\Variable{op}\;\Variable{v}_2\;
							\rightarrow^e\;
				\boldsymbol{\delta}(\Variable{op}, \nu_1, \nu_2)
				}[]
			$$

			$$\setnamespace{0.5em}
				\inference{\Variable{op} \in \{ \textbf{+}, \textbf{-}, \textbf{*}, \textbf{/},
				                                \boldsymbol{\hat{\null}}\;,
				                                \textbf{\%}\}\\
					\Variable{v}_1\;\notin\;\Nonterminal{number}\;
					                \vee\;
					\Variable{v}_2\;\notin\;\Nonterminal{number}\\
					(\boldsymbol{\delta}(\Keyword{tonumber},\;\Variable{v}_1, 10)\;
					                \notin\;\Nonterminal{number}\;
						            \vee\\
					\boldsymbol{\delta}(\Keyword{tonumber},\;\Variable{v}_2, 10)\;\notin\;\Nonterminal{number})
				}
				{\Variable{v}_1\;\Variable{op}\;\Variable{v}_2\;
							\rightarrow^e\;
						\LabelCons{\Variable{v}_1\;\Variable{op}\;\Variable{v}_2}
						          {\Label{ArithWO}}}
	            []$$

\caption{Semantics of stateless expressions.}
\flabel{semantics_simple_expressions}
\end{figure}

We extend the {\em stateless} subset presented in \sref{formalism_used} with
while loops, breaks, composition of statements, and numbers and strings with their
corresponding operations  (\fref{syntax_simple}).

Correspondingly, we extend the
relation $\rightarrow^{\textsf{s}}$ with
the semantics of the new statements (\fref{semantics_simple_statements}).
First, a while loop begins by wrapping the whole loop in a \Keyword{Break}
label, changing also the name from \Keyword{while} to \Keyword{\$iter}. The
purpose of the label is to mark the point in which a \Keyword{break} should
continue the execution, and the renaming is necessary to avoid repeatedly
unfolding a while and piling up labels. Labels and \Keyword{\$iter} are new
run-time statements (\fref{while_runtime}).
Then, a loop marked with \Keyword{\$iter} is unfolded as usual, using the
conditional to check the guard and perform a new iteration.

In a composition of
two statements, when the one on the left is a skip (\Keyword{;}), we continue
with the second.
More interestingly, when the execution finds a \Keyword{break} inside a
(labeled) block, the whole code is replaced with a skip, to signal the execution
of the break has exited. This is achieved by using a new evaluation context
$E_\textsf{lf}$, which represents a program in which no other labeled term occur
(its definition is elided for brevity). By not having other labels, we
know the one surrounding this context is the one we have to break. The last rule
removes the label once the execution of a loop reached skip.

Having defined the semantics for statements, we turn our attention to
expressions (\fref{semantics_simple_expressions}). For brevity we focus only on
arithmetic operators, but similar rules exists for strings.  The first rule
state that, if operands $\Variable{v}_1$ and $\Variable{v}_2$ are both numbers,
and the operation is relevant to numbers, we delegate the result to the $\delta$
function already introduced in \ref{sec:formalism_used}.  The second rule covers
the case where one or both of the operands are not numbers, but can be coerced
into a number by the external function $\Keyword{tonumber}$. In that case, we coerce the
operands and do the operation.  There is similar rule for concatenation, elided
for brevity, when one of the operands is a string and the other a number.
Finally, the last rule applies when the operands cannot be coerced into
numbers. In this case we label the expression with \Label{ArithWO} (some labels are
listed in \fref{runtime_constructions_exps_no_store}, where \Label{WO} stands for Wrong Operands), to signal the error. At
this point, execution is stuck here, but in \sref{metatables} we show how the
metatable mechanism handles this erroneous situation.

\subsection{Imperative Lua}
\label{sec:stat_expr_store}

\begin{figure}
{
    \begin{tabbing}
    	\Nonterminal{v} ::= \= \ldots{} \sep \Keyword{function} \Nonterminal{l} \Terminal{(} \Nonterminal{x} \Terminal{,} \ldots{}
    	\Terminal{)} \Nonterminal{s} \Keyword{end}\\
    	\>\sep \Keyword{function} \Nonterminal{l} \Terminal{(} \Nonterminal{x} \Terminal{,} \ldots{}
    	\Terminal{, ...)} \Nonterminal{s} \Keyword{end}\\
    	
        \Nonterminal{s} ::= \=\;\;\ldots{} \sep \Nonterminal{e}
		                                        \Terminal{(} \Nonterminal{e} \Terminal{,} \ldots{} \Terminal{)}
					        \sep \Nonterminal{e} \Terminal{:} \Nonterminal{x}
					                            \Terminal{(} \Nonterminal{e} \Terminal{,} \ldots{} \Terminal{)}
					      \sep \Keyword{return} \Nonterminal{e}\\
      \> \sep \Keyword{local} 
      \;\Nonterminal{x}, 
      \ldots{}\,\Terminal{=}\,
      \Nonterminal{e}, \ldots{}  
      \Keyword{in}\,
      \Nonterminal{s}
      \Keyword{end} 
      
      \sep\Nonterminal{var} \Terminal{,} \ldots{}
      \Terminal{=} \Nonterminal{e} \Terminal{,} \ldots{} \\
			                     
        \Nonterminal{var} ::= \= \Nonterminal{x} 
	                  \sep \Nonterminal{e} \Terminal{[} \Nonterminal{e} \Terminal{]}\\
		
		\Nonterminal{e} ::= \= \ldots{} \sep \Terminal{(}\! \Nonterminal{e}\! \Terminal{)} \sep
 \Terminal{\{}\! \Nonterminal{field}
		                                                \Terminal{,} \ldots{}
		                                     \!\Terminal{\}}
				  \sep \Nonterminal{e}\! \Terminal{(}\! \Nonterminal{e}\! \Terminal{,} \ldots{}\! \Terminal{)}
                                  \sep \Nonterminal{e}\! \Terminal{:} \Nonterminal{x} \Terminal{(}\! \Nonterminal{e} \Terminal{,} \ldots{}\! \Terminal{)}\\
			     
        \Nonterminal{field} ::= \Nonterminal{e}
                     \sep \Terminal{[} \Nonterminal{e} \Terminal{] =} \Nonterminal{e}
        \end{tabbing}}
\caption{Syntax of the remaining imperative subset.}
\flabel{syntax_s_e_store}
\end{figure}

\begin{figure}
{
    \begin{tabbing}
        \Nonterminal{v} ::= \= \ldots{} \sep \Nonterminal{objr}\\

        \Nonterminal{e} ::= \= \ldots{} \sep  \Mterminal{<} \Nonterminal{e} \Terminal{,} ... \Mterminal{>} \\
										  
		\Nonterminal{label} ::= ... \sep \Label{Return} \sep \Label{Index} \sep \Label{NewIndex} \sep \Label{WFunCall}
		\end{tabbing}}
\caption{Store-related run-time terms.}
\flabel{store_runtime_constructions}
\end{figure}

\begin{figure}
	$$\setnamespace{0.5em}
			\inference{\mathsf{\boldsymbol{\delta}(rawget, 
					   								\Variable{objr}, 
					   								\Variable{v}_1,
					   								\theta_1) \neq \Keyword{nil}}\\
					   \mathsf{\theta_2 = \mathsf{\boldsymbol{\delta}(rawset, 
					   								\Variable{objr}, 
					   								\Variable{v}_1,
					   								\Variable{v}_2,
					   								\theta_1)}}}
					{\mathsf{\theta_1\;\boldsymbol{:}\;
							\Variable{objr}\;\Terminal{[}\Variable{v}_1 \Terminal{]}\;
							\Terminal{=}\;\Variable{v}_2\;
								\rightarrow^{{\sf s\_\theta}}\;
							\theta_2\;\boldsymbol{:}\;\Terminal{;}}}
			[] $$\\

	$$\setnamespace{0.5em}
			\inference{\mathsf{\boldsymbol{\delta}(rawget, 
					   								\Variable{objr}, 
					   								\Variable{v}_1,
					   								\theta) = \Keyword{nil}}}
					{\mathsf{\theta\;\boldsymbol{:}\;
							\Variable{objr}\;\Terminal{[}\Variable{v}_1 \Terminal{]}\;
							\Terminal{=}\;\Variable{v}_2\;
								\rightarrow^{{\sf s\_\theta}}\;
							\theta\;\boldsymbol{:}\;
							\LabelCons{\Variable{objr}\;\Terminal{[}\Variable{v}_1 \Terminal{]}\;
									\Terminal{=}\;\Variable{v}_2 }
									{\Label{NewIndex}}}}
			[] $$\\
			
	$$\setnamespace{0.5em}
			\inference{\mathsf{\boldsymbol{\delta}(type, \Variable{v}_1) \neq "table"}}
					{\mathsf{\theta\;\boldsymbol{:}\;
							\Variable{v}_1\;\Terminal{[}\Variable{v}_2 \Terminal{]}\;
							\Terminal{=}\;\Variable{v}_3\;
								\rightarrow^{{\sf s\_\theta}}\;
							\theta\;\boldsymbol{:}\;
							\LabelCons{ 
							\Variable{v}_1\;\Terminal{[}\Variable{v}_2 \Terminal{]}\;
									\Terminal{=}\;\Variable{v}_3}
						    {\Label{NewIndex}}}}
			[] $$\\
\caption{Field update.}
\flabel{semantics_statements_theta}
\end{figure}

\begin{figure}
	
	$$\setnamespace{0.5em}
		\inference{\mathsf{\Variable{v}_2 \;=\; 
							\boldsymbol{\delta}(rawget,\;
														\Variable{objr},\;
														\Variable{v}_1,\;\theta)}\;\;
					\mathsf{\Variable{v}_2 \neq \Keyword{nil}}}	
		{\mathsf{\theta\;\textbf{:}\;\Variable{objr}\;\textbf{[}\;\Variable{v}_1\;\textbf{]}\;
			\rightarrow^{\mathsf{e\_\theta}}\;\theta\;\textbf{:}\;
			\Variable{v}_2}}
			[]$$
			
			$$
			\setnamespace{0.5em}
		\inference{\mathsf{ 
							\boldsymbol{\delta}(rawget,\;
														\Variable{objr},\;
														\Variable{v},\;\theta)\;=\;\Keyword{nil}}}	
		           {\mathsf{\theta\;\textbf{:}\;\Variable{objr}\;
		                    \Terminal{[}\;\Variable{v}\;\Terminal{]}\;
			\rightarrow^{\mathsf{e\_\theta}}
			    \;\theta\;\textbf{:}\;
			    \LabelCons{\Variable{objr}\;\Terminal{[}\;\Variable{v}\;\Terminal{]}}
			              {\Label{Index}}}}[] $$
	
	$$\setnamespace{0.5em}
	\inference{\mathsf{\boldsymbol{\delta}(type, \Variable{v}_1) \neq ``table"}}
	          {\mathsf{\theta\;\textbf{:}\;\Variable{v}_1\;
	                                      \Terminal{[}\;\Variable{v}_2\;\Terminal{]}\;
	                                      \rightarrow^{e\_\theta}
	                   \;\theta\;\textbf{:}\;
	                                      \LabelCons{\Variable{v}_1\;\Terminal{[}\;\Variable{v}_2\;\Terminal{]}}
	                                                {\Label{Index}}}}[] $$
\caption{Field indexing.}
\flabel{expr_theta_stores_tab_index}
\end{figure}

\begin{figure}
	$$\setnamespace{0.5em}
		\inference{\mathsf{\forall\;1 \leq i, 
					\Variable{field}_i = \Variable{v}
					\vee
					\Variable{field}_i = \textbf{[}\;\Variable{v}\;\textbf{]}\;\textbf{=}\;\Variable{v}'}\\
					\mathsf{\theta_2 = \textbf{(}\;\Variable{objr},
					\;\boldsymbol{<}\;addkeys(\textbf{\{}\Variable{field}_1 \textbf{,}\;... \textbf{\}})\;												\textbf{,}
									\;\nil\;\boldsymbol{>}\textbf{)},\;\theta_1}}
	{\mathsf{\theta_1\;\textbf{:}\;\textbf{\{} \Variable{field}_1 \textbf{,}\;... \textbf{\}} 
	\;\rightarrow^{\mathsf{e\_\theta}}\;\theta_2\;\textbf{:}\;\Variable{objr}}}[] $$
\caption{Object creation.}
\flabel{expr_theta_stores_obj_creat}
\end{figure}

\begin{figure}[!htbp]
	$$\setnamespace{0.5em}
		\inference
		{\mathsf{\sigma' = (\Variable{r}_1, \Variable{v'}_1),...,
		                   (\Variable{r}_{\textsf{n}}, \Variable{v'}_n), \sigma}\\
        \mathsf{i \leq m \Rightarrow \Variable{v'}_i = \Variable{v}_i}\;\;\;\;
                    \mathsf{i > m \Rightarrow \Variable{v'}_i = \Keyword{nil}}}
		{\begin{aligned}\mathsf{\sigma\;\textbf{:}\;(\Keyword{function}}&\mathsf{\;\Variable{l}\;
				(\Variable{x}_1,...,\Variable{x}_n)\;
				\Variable{s}\;
				\Keyword{end})\;}\mathsf{
		                        (\Variable{v}_1},...,\Variable{v}_{\textsf{m}})\; 
								    \rightarrow^{\mathsf{funcall}}\\
		        & \sigma'\;\textbf{:}\;
		                        \LabelCons{\Variable{s}\;
		                            [\Variable{x}_1 \backslash \Variable{r}_1,...,
		                             \Variable{x}_{\textsf{n}} \backslash \Variable{r}_{\textsf{n}}]}
		                             {\Label{Return}}\end{aligned}}
        $$
        
    $$\setnamespace{0.5em}
		\inference
		{\mathsf{\sigma' = (\Variable{r}_1, \Variable{v}_1),...,
		                  (\Variable{r}_{\textsf{n}}, \Variable{v}_n), \sigma}\\
        \mathsf{i \leq m \Rightarrow} \mathsf{\Variable{v'}_i = \Variable{v}_i\;\;\;\;
                    i > m \Rightarrow \Variable{v'}_i = \Keyword{nil}}\\
		\mathsf{\Variable{tuple} = \;\boldsymbol{<} \Variable{v}_{n + 1},...,
		                                          \Variable{v}_{m} \boldsymbol{>}}}
		{\begin{aligned}\mathsf{\sigma\;\textbf{:}\;} & \mathsf{(\Keyword{function}\;\Variable{l}\;
				(\Variable{x}_1,...,\Variable{x}_n, \textbf{...})\;
				\Variable{s}\;
				\Keyword{end})\;(\Variable{v}_1},...,
		                                                                   \Variable{v}_{\textsf{m}})\;
		         \\&\rightarrow^{\mathsf{funcall}}
		        \sigma'\;\textbf{:}\;
		                \LabelCons{\Variable{s}\;
		                [\Variable{x}_1 \backslash \Variable{r}_1,...,
		                 \Variable{x}_{\textsf{n}} \backslash \Variable{r}_{\textsf{n}}, 
		                 \textbf{...}\backslash\Variable{tuple}]}
		                 {\Label{Return}}\end{aligned}}
	$$\\
	
	$$\setnamespace{0.5em}
		\inference
		{\mathsf{\boldsymbol{\delta}(type, \Variable{v}) \neq \dbq{function}}}
		{\mathsf{\sigma\;\textbf{:}} \mathsf{\;\Variable{v}\;
		                                        (\Variable{v}_1,...,\Variable{v}_{\textsf{n}})\;
				\rightarrow^{\mathsf{funcall}}\;} \mathsf{
		        \sigma\;\textbf{:}\;
		        \LabelCons{\Variable{v}\;(\Variable{v}_1,...,\Variable{v}_{\textsf{n}})\;}
		                  {\Label{WFunCall}}}}
	$$\\
	
	$$\setnamespace{0.5em}
\begin{aligned}
	    	\mathsf{\sigma\;\textbf{:}\;
	         \Variable{v}\Terminal{:}} & \mathsf{\Variable{name}\;
			 \Terminal{(}\Variable{e}_1\Terminal{,} ...\Terminal{,}\Variable{e}_n \Terminal{)}\;
			 \rightarrow^{\mathsf{funcall}}}\mathsf{
			 \sigma\;\textbf{:}\;
			 \Variable{v}\Terminal{[``}\Variable{name} \Terminal{"} 
			 \Terminal{] (} \Variable{v}\Terminal{,}\Variable{e}_1\Terminal{,} ...\Terminal{,} 
			 \Variable{e}_n\Terminal{)}
			 }
\end{aligned}
	$$
	
\caption{Function and method calls.}
\flabel{functioncall}
\end{figure}

\begin{figure}
\begin{tabular}{c}
						   	$\mathsf{\sigma\;\textbf{:}\;\LabelCons{\Terminal{;}}
						                      {\Label{Return}}
									    \rightarrow ^{\textsf{funcall}}\;
									\sigma\;\textbf{:}\;\Terminal{;}}$\\

						$\mathsf{
						    \sigma\;\textbf{:}\;\LabelCons{\Variable{E}_\textsf{lf} [\![\;
									   \Keyword{return}\;< \Variable{v} , ... > \;]\!]}
									   {\Label{Return}}
							\;\rightarrow ^{\textsf{funcall}}}$\\
						 $\mathsf{\sigma\;\textbf{:}\;< \Variable{v} , ... >}$\\

						$\mathsf{
						    \sigma\;\textbf{:}\;\LabelCons{\Variable{E}_\textsf{lf} [\![\;
									   \Keyword{return}\;< \Variable{v} , ... > \;]\!]}
									   {\Label{Break}}
							\;\rightarrow ^{\textsf{funcall}}}$\\
							$\mathsf{\sigma\;\textbf{:}\;\Keyword{return}\;< \Variable{v} , ... >}$
\end{tabular}
\caption{Semantics of \Keyword{return}.}
\flabel{return}
\end{figure}

The {\em imperative} subset is made up of functions, function
application, tables, field indexing, and field update. Despite being
values, Lua functions are in the imperative subset because parameters
are mutable variables, so they are allocated in the $\sigma$ store.
Tables are mutable objects, and we allocate them in a separate
store, denoted with $\theta$. Object references, the domain of $\theta$,
are considered values, so are in the image of $\sigma$. We ask form them
to satisfy the same properties as asked for references $\sigma$,
together with the possibility of distinguish syntactically between
each kind of reference. The image of $\theta$ only contains tables.

Functions are labeled so each function in the source program
has a unique label \Variable{l}. How labels are represented is not important; as long as they are comparable.
This reproduces the correct semantics of function equality in Lua,
where two identical functions are not equal if they are defined in
different parts of the source file, as shown in the following
interaction with the reference interpreter:
\begin{lstlisting}[numbers=none]
> f = function() end
> g = function() end
> print(f == g)
false
\end{lstlisting}

A source function may evaluate to different values during
the evaluation of the program, due to different substitutions
of their free variables. Our use of substitution and references
means that we do not need to have explicit {\em closures},
a function definition is itself a closure once the focus
of evaluation has reached it.

\fref{store_runtime_constructions} adds \emph{tuples}, used for
returning multiple values from function application, and for
functions that can handle a variable number of arguments ({\em vararg}
functions) through the \Terminal{...} {\em vararg} operator. Wrapping
an expression in parenthesis has a semantic effect in Lua: if the
expression evaluates to a tuple the parenthesis discards all but
the first value of the tuple (if the tuple is empty the parenthesized
expression evaluates to \nil).

Besides being ``truncated" to their first value, these tuples
can also be concatenated
with another tuple, depending on their syntactical place in the program:
in an expression list $e_1,\ldots,e_n$ the tuples of expressions
$e_1$ to $e_{n-1}$ evaluate to their first value, or \nil{} for the
empty tuple (the same behavior as parenthesized expressions). These $n-1$ values then form a tuple of their own, which is
concatenated with the tuple of $e_n$. Semantically, this is done
through reductions between tuples that ``flattens" tuples of tuples
until reaching a tuple where none of the values are another tuple.

The new statements also include multiple variable definition and assignment, which generalizes the single-variable versions introduced in
\sref{formalism_used}. The reduction rules for these statements are not 
shown here for reasons of brevity, but are a straightforward extension of the simpler versions: in case of multiple assignment, the evaluation contexts assure that all lvalues are evaluated before rvalues,
and the tuples for both sides are flattened, then lvalues are paired with their corresponding rvalue, with
any lvalues that do not have a corresponding rvalue paired with \nil.

\fref{semantics_statements_theta} describes assignment to table fields.
It uses some services modeled by the $\delta$ function: $\mathsf{\delta(type,v)}$ is the type of the value; 
$\delta(\mathsf{rawget},\Variable{objr},\Variable{v},\theta)$ is primitive table indexing,
yielding either the value associated with $\Variable{v}$ in $\theta(\Variable{objr})$
or \Keyword{nil} if there is no associated value;
$\delta(\mathsf{rawset},\Variable{objr},\Variable{v}_k,\Variable{v},\theta)$ is primitive table update,
yielding a new $\theta$ where the table referenced by $\Variable{objr}$ associates \Variable{v} with value $\Variable{v}_k$.

The rules show field update under 3 different circumstances: when 
the operation is made over an actual table with an existing key;
when the operation is made over an actual table but with an unknown key;
and when the assignment is carried over a non-table value. The last
two cases just tag the expression with \Label{NewIndex}, which will be
handled by the metatable mechanism explained in Section \ref{sec:metatables}. Field access (\fref{expr_theta_stores_tab_index}) have similar rules, but
tagging exceptional situations with \Label{Index}.

\fref{expr_theta_stores_obj_creat} provides meaning to
table constructors. Its complete semantics actually depends upon the 
meta-function \textsf{addkeys}, which adds absent keys in the constructor
(see \fref{syntax_s_e_store} for the syntax of table constructors).
It works by supplying consecutive natural numbers as keys, starting with 1.

\fref{functioncall} shows function and method application, described with a new relation $\rightarrow^{\mathsf{funcall}}$. Formal
parameters are mutable variables, so a fresh reference is allocated
for each parameter. The first rule
covers all the cases involving the application of a non-vararg function:
when it is applied to the same number of arguments as formal parameters,
when it is applied to fewer arguments, with unpaired parameters receiving
\Keyword{nil}, and when it is applied to more arguments, with extra
arguments silently ignored. Similar to what we did with while loops in the previous section, we also label the body with \Label{Return},
to indicate the point to which a \Keyword{return} statement must
go. It is, roughly speaking, the syntactic equivalent
to the return address saved in an activation frame.

Returning to \fref{functioncall}, the second rule shows the case of a 
vararg function call: the difference just resides on what is done
with surplus arguments: in this case, they are put into a tuple expression, which replaces the vararg expression (\Keyword{...}) in the body of the function. 

The third rule has to do with one of the exceptional situations that can
be managed by the metatable mechanism: a function call over a non-function
value. Again, at this point we just label
the whole expression with a tag that indicates what happened.  The last rule
 shows how the method invocation is translated into a table look-up, with the
object being injected as the first argument of the function.

\fref{return} shows the semantics of the \Keyword{return} statement as well as
implicitly returning by reaching the end of the function. The ideas used in this
rules are analogous to the ones expressed when defining the semantics of the
\Keyword{break} statement, in Section \ref{sec:stat_expr_no_store}.

\subsection{Built-in services}
\label{sec:built_in_sec}

 \begin{figure}
 		$$\setnamespace{0.5em}
 		\inference{\mathsf{l \in \{ type, assert, error, pcall, select, ... \}}}
 		{\mathsf{\theta\;\Keyword{:}\;\Keyword{\$builtIn}\;l\;\Terminal{(} 
 					\Variable{v}_1,...,\Variable{v}_n \Terminal{)}
 		\rightarrow^{{\sf builtIn}}
 		\theta\;\Keyword{:}\;\;\boldsymbol{\delta} (l,\;\Variable{v}_1,...,
 		                                            \Variable{v}_n)}}[] $$ \\

 		$$\setnamespace{0.5em}
 		\inference{\mathsf{l \in \{ ipairs, next, pairs, getmetatable, ... \}}}
 		{\mathsf{\theta\;\textbf{:}\;\Keyword{\$builtIn}\;l\;\Terminal{(} 
 					\Variable{v}_1,...,\Variable{v}_n \Terminal{)}
 		\rightarrow^{{\sf builtIn}}
 		\theta\;\textbf{:}\;\boldsymbol{\delta} (l,
 												\Variable{v}_1,...,
 												\Variable{v}_n, 
 												\theta)}}[] $$ \\

 		$$\setnamespace{0.5em}
 		\inference{\mathsf{l \in \{ rawset, setmetatable \}}\\
 		           \mathsf{\theta_2 = \boldsymbol{\delta} (l, 
 										\Variable{v}_1,..., 
 										\Variable{v}_n,
 										\theta_1)}}
 		{\mathsf{\theta_1\;\textbf{:}\;
 				\Keyword{\$builtIn}\;l\;\Terminal{(}
 								\Variable{v}_1,...,
 								\Variable{v}_n \Terminal{)}
 		\rightarrow^{{\sf builtIn}}
 		\theta_2\;\textbf{:}\;\Variable{v}_1}}[] $$

 \caption{Interface with the $\delta$ function.}
 \flabel{built_in_rules}
 \end{figure}

 In Lua, built-in services offered by Lua's standard library are stored in the
 execution environment, a table named $\mathsf{\_ENV}$, where the keys are the
 names of the services and the values are their definitions. For instance, when we
 access the table field named ``type", we access the function that given an
 element provides its type (as a string):
\begin{lstlisting}[numbers=none]
> print(type({}))
table
\end{lstlisting}
(Remember from \sref{langintro}: using an identifier not in scope is equal to accessing $\mathsf{\_ENV}$.) We can override its definition and obtain a different behavior:
\begin{lstlisting}[numbers=none]
> type = function () return 'not a type' end
> print(type({}))
not a type
\end{lstlisting}
However, the original $\mathsf{type}$ function is still accessible from other
services in the library. We can see this when we call $\mathsf{next}$, the
function that iterates over the fields in a table:
\begin{lstlisting}[numbers=none]
> next(1)
stdin:1: bad argument (table expected, got number)
\end{lstlisting}

In order to model this behavior, prior to the execution of a program the
$\mathsf{\_ENV}$ table must be populated with the functions from the standard
library. But these functions are just wrappers for a special (run-time)
expression \Keyword{\$builtIn}. When evaluated, this expression calls the
$\delta$ function with the actual definition of the function. Built-in services,
like $\mathsf{next}$, might call other services through the \Keyword{\$builtIn}
term instead of ordinary function application, effectively reproducing the early
binding that is required.

While it might sound a bit intricate, this design gives the formalization
several desirable properties: compliance with the semantics as defined in the
reference manual and the reference
interpreter, and a modular way of tackling the formalization of built-in
services. And at the level of the mechanization, it allows us to experiment and
test against different implementations of these services with minimal changes in
the rest of the formalization.

\fref{built_in_rules} gives the semantics of \Keyword{\$builtIn}
using three rules, corresponding to three different kinds
of services: services that do not operate on tables, so
do not need to access the object store $\theta$, services that read
from tables, and services that update existing tables, yielding
a new $\theta$. The antecedents of the first two rules show
just some of the services that are in each category; the actual
list of services includes almost all the built-in basic functions of
the Lua language, together with services from the libraries
\textsf{math}, \textsf{string} and \textsf{table}.

\begin{figure}
  \begin{tabbing}
    $\mathsf{\boldsymbol{\delta}(pairs,}$\= $\mathsf{\Variable{objr}, \theta)}$ =
			\lstset{emph={obj, any}, emphstyle=\itshape}
\begin{lstlisting}[numbers=none]
(function $getIter ()
   local v1, v2, v3 = h(objr) in
     return < v1, v2, v3 >
 end)()
\end{lstlisting}\\
 \> where $\Variable{h} =\mathsf{indexmetatable}($\= $\Variable{objr}, ``\mathsf{\_\_pairs}", \theta)$\\
 \> and $\Variable{h} \neq \Keyword{nil}$\\

	$\mathsf{\boldsymbol{\delta}(pairs,}$\= $\mathsf{\Variable{objr}, \theta)}$ =
				\lstset{emph={obj, any}, emphstyle=\itshape}
                \begin{lstlisting}[numbers=none]
< function $next (table, index)
    return $builtIn next(table, index)
  end, objr, nil>
                \end{lstlisting}\\
                \> if $\mathsf{indexmetatable}(\Variable{objr},``\mathsf{\_\_pairs}",\theta) = \Keyword{nil}$\\[.4em]

    $\mathsf{\boldsymbol{\delta}(pairs,}$\= $\mathsf{\Variable{v}, \theta)}$  =
		$\Keyword{\$builtIn}~\mathsf{error}(\Variable{msg}\,..\, \Keyword{\$builtIn}~\mathsf{type}(\Variable{v}))$\\
		 \> if $\mathsf{\boldsymbol{\delta} (type, \Variable{v}) \neq ``table"}$\\
                 \> where \Variable{msg} = ``table\;expected,\;got\;"
	\end{tabbing}
\caption{Basic functions of the
standard library: pairs.}
\flabel{basic_functions_pairs}
\end{figure}

The $\delta$ function defines, in a denotational way, the actual fundamental
details of the semantics of the built-in services and the primitive operators of
the language. In the rest of this section we discuss an interesting example: the
{\sf pairs} built-in function (\fref{basic_functions_pairs}).

The built-in service \textsf{pairs} is used to iterate a table using a
\Keyword{for} loop. It must return three values: an iterator function, the
object to index, and the first index. According to the equations in the figure,
there are three different scenarios: In the first case, when the table
\Variable{objr} has a custom handler \Variable{h} in the 
{\sf \_\_pairs} key of its metatable, calls this handler to
get the iterator triplet.
The metafunction \textsf{indexmetatable} queries the metatable
(metatables are discussed further in \sref{metatables}).
Also note that we let $\delta$ yield not only values but any
valid expression.

It might look odd why we create a function whose body calls
\Variable{h}, instead of directly returning it. The reason is twofold: First,
according to the manual, we must only return the first three values returned by
the indexed function. This is achieved by creating three variables, one for each
value, and return only those. If there are more values, they are
discarded. Second, since \Keyword{local} and \Keyword{return} are not valid
\emph{expressions}, and in this case, $\boldsymbol{\delta}$ must return an
expression (not necessarily a value!), we wrap this code in a closure.

In the second case, when the table has no metatable or no handler for
{\sf \_\_pairs}, the reference manual indicates
that \textsf{pairs(t)} returns ``\textit{the \texttt{next} function, the table
\texttt{t}, and \Keyword{nil}}". 
This case models this behavior by wrapping a call to
\Keyword{\$builtIn} {\sf next}, as mentioned earlier in this section.
The label of this function guarantees that it will be the same
function that is bound to {\sf next} in our initial environment.

The third and final case of {\sf pairs} constructs an expression that
will assemble an error message and then throw an error using the {\sf error}
built-in primitive. As with {\sf next}, we cannot look-up the {\sf error}
built-in function in the environment, as it could have been rebound by
the programmer.

Before concluding this section, we note that we let the interpretation function
define the meaning of every primitive operator and library service using a
denotational approach, following the tradition of Landin's ISWIM language~\cite{700}. Several of
these primitives could also be given operational semantics, however, we decide
to stick to ISWIM's philosophy, which prioritizes cohesion, modularity, and
expressivity.

\subsection{Metatables}
\label{sec:metatables}
\begin{figure*}
\begin{tabular}{c}
    $$\setnamespace{0.5em}
	 	\inference{\mathsf{\Variable{v}_3 = getbinhandler(\Variable{v}_1, \Variable{v}_2,
	 	                                    binopeventkey(\Variable{op}), \theta)}\;\;\;\;\;\;
				   \mathsf{\Variable{v}_3 \neq \nil}
					        }
		{\mathsf{\theta\;\textbf{:}\;
		        \LabelCons{\Variable{v}_1\;\Variable{op}\;\Variable{v}_2}{\Label{ArithWO}}\;
				\rightarrow^{\textsf{e\_metatable}}
				\theta\;\textbf{:}\;\;
				\Variable{v}_3\;\Terminal{(}\Variable{v}_1 \Terminal{,}\;
				                            \Variable{v}_2 \Terminal{)}}}
        []$$\\\\

    $$\setnamespace{0.5em}
    \inference{\mathsf{getbinhandler(\Variable{v}_1, \Variable{v}_2,
    		binopeventkey(\Variable{op}), \theta)
    	= \nil\;\;\;\;\;\;\mathsf{\Variable{t}_1 = \boldsymbol{\delta}(type, \Variable{v}_1)}\;\;\;\;\;\;\mathsf{\Variable{t}_2 = \boldsymbol{\delta}(type, \Variable{v}_2)}}
    }
    {\mathsf{\theta\;\textbf{:}\;
    		\LabelCons{\Variable{v}_1\;\Variable{op}\;\Variable{v}_2}{\Label{ArithWO}}\;
    		\rightarrow^{\textsf{e\_metatable}}
    		\theta\;\textbf{:}\;\Keyword{\$builtIn}\;
    		error\;(\;\textsf{\textbf{\#errmessage}}(\Label{ArithWO},
    		\Variable{t}_1,\Variable{t}_2)\;)}}
    []$$
\end{tabular}\\[.4em]
\caption{Metatable mechanism for arithmetic binary expressions.}
\flabel{metatables_expressions_arith_string}
\end{figure*}

\begin{figure*}
\begin{tabular}{c}
	    $$\setnamespace{0.5em}
			\inference{\mathsf{\Variable{v}_4 = indexmetatable(\Variable{v}_1,
			                                              ``\_\_newindex",
																\theta)}\;\;\;\;\;\;
						\mathsf{\boldsymbol{\delta}(type, \Variable{v}_4) = ``function"}}
						{\mathsf{\theta\; \textbf{:}
								\;\LabelCons{\Variable{v}_1\; \Terminal{[} \Variable{v}_2
								\Terminal{]}\; \Terminal{=}\; \Variable{v}_3}
								{\Label{NewIndex}}
 									\rightarrow^{\textsf{s\_metatable}}
								\;\theta\; \textbf{:}\;
									\Variable{v}_4\; \Terminal{(} \Variable{v}_1,
																\Variable{v}_2,
																\Variable{v}_3
													\Terminal{)}}}
			[] $$\\\\

	    $$\setnamespace{0.5em}
	    \inference{\mathsf{\Variable{v}_4 = indexmetatable(\Variable{v}_1,
	    		``\_\_newindex",
	    		\theta)}\;\;\;\;\;\;\Variable{v}_4 \neq \nil\;\;\;\;\;\;
	    	\mathsf{\boldsymbol{\delta}(type, \Variable{v}_4) \neq ``function"}}
	    {\mathsf{\theta\; \textbf{:}
	    		\;\LabelCons{\Variable{v}_1\; \Terminal{[} \Variable{v}_2
	    			\Terminal{]}\; \Terminal{=}\; \Variable{v}_3}
	    		{\Label{NewIndex}}
	    		\rightarrow^{\textsf{s\_metatable}}
	    		\;\theta\; \textbf{:}\;
	    		\Variable{v}_4\; \Terminal{[}
	    		\Variable{v}_2\Terminal{]} \;\Terminal{=}\;
	    		\Variable{v}_3}}
	    [] $$\\\\

		$$\setnamespace{0.5em}
		\inference{\mathsf{indexmetatable(objr,``\_\_newindex",
				\theta_1) = \nil}\;\;\;\;\;\;
			\mathsf{\theta_2 = \boldsymbol{\delta}(rawset, objr, \Variable{v}_1, \Variable{v}_2, \theta_1)}}
		{\mathsf{\theta_1\; \textbf{:}
				\;\LabelCons{objr\;\Terminal{[}
					\Variable{v}_1 \Terminal{]}\; \Terminal{=}\;
					\Variable{v}_2}{\Label{NewIndex}}
				\rightarrow^{\textsf{s\_metatable}}\;\theta_2\;\textbf{:}\;;}}
		[] $$\\\\
		
		$$\setnamespace{0.5em}
			\inference{\mathsf{indexmetatable(\Variable{v}_1,
											``\_\_newindex",
											\theta) = \nil}\;\;\;\;\;\;
						\mathsf{\Variable{t} = \boldsymbol{\delta}(type, \Variable{v}_1)}\;\;\;\;\;\;\mathsf{t \neq ``table"}}
					{\mathsf{\theta\; \textbf{:}
								\;\LabelCons{\Variable{v}_1\;\Terminal{[}
								             \Variable{v}_2 \Terminal{]}\; \Terminal{=}\;
								             \Variable{v}_3}{\Label{NewIndex}}
									\rightarrow^{\textsf{s\_metatable}}\;
							\theta\; \textbf{:}\;
								\Keyword{\$builtIn}\; error\; \Terminal{(}\;
																\textsf{\textbf{\#errmessage}}(
																			\Label{NewIndex},
																			\Variable{t})\;
																\Terminal{)}}}
			[] $$\\
\end{tabular}\\[.4em]
\caption{Metatable mechanism for field update.}
\flabel{metatables_stats}
\end{figure*}

The most notable feature of Lua is its metaprogramming mechanism,
{\em metatables}, that lets the programmer adapt the language to
specific domains. With metatables Lua can maintain its original
design decision to ``\textit{keep the language simple and 
small}"\cite{evlua}, while still being able to cope with
a variety of programming concepts\footnote{See section
``\textsf{Code Structure / Programming Paradigms}" at
\url{lua-users.org/wiki/LuaDirectory}}.

Briefly, metatables let the programmer specify {\em fallbacks} for
certain operations: arithmetic over non-numeric values,
concatenation over non-string values,
equality between objects that do not have the same identity,
application over values that are not functions, indexing or
updating a field over a value that is not a table, etc.

Metatables are plain tables, and the fallbacks that a particular
metatable supports are typically functions associated with a
unique string key for each operation (e.g. {\sf \_\_add} for
the fallback to the plus operator, or {\sf \_\_newindex} for
the fallback to field update). Lua libraries are free to
extend this mechanism with their own fallbacks (like the {\sf pairs}
built-in function of the previous section, which can look up {\sf \_\_pairs} in the metatable, if it exists).
Each Lua table can have its own metatable, while values of other
types share a single metatable for each type.

We have shown in previous sections that regular semantics
of operations just labels the expression or statement involved
when it reaches a case where a fallback in a metatable could
be used. This approach simplifies the regular semantics,
and improves the modularity of the formalization.
The relations $\rightarrow^{\textsf{e\_metatable}}$ 
and $\rightarrow^{\textsf{s\_metatable}}$ that we show
in this section take these labeled terms and act accordingly.

\fref{metatables_expressions_arith_string} shows how $\rightarrow^{\textsf{e\_metatable}}$
resolves arithmetic operations over operands
of unexpected type, a condition labeled with \Label{ArithWO}.
The metafunction \textsf{getbinhandler} is analogous to those described in the Lua reference
manual: 
it looks for a handler first in the metatable of the left operand $\Variable{v}_1$, then
the metatable of the right one $\Variable{v}_2$, 
by looking into the corresponding field in those metatables.
We also abstract the mapping between binary operators and their
metatable keys with the metafunction \textsf{binopeventkey}.

Looking up a fallback in a metatable is guaranteed to either return the 
fallback or return \Keyword{nil} because of two invariants: a metatable
is always a table, and the metatable of a metatable, it if exists, is
ignored for this look-up. This means that abstracting this look-up
with a metafunction does not compromise the small-step nature
of our semantics.

The first rule of \fref{metatables_expressions_arith_string}
shows how the operation is rewritten as an application
of the handler on the two operands as arguments. If the handler is
not a function this may trigger yet another fallback.
The second rule shows what happens when no handler is found:
an error is thrown using the {\sf error}
built-in service. We also abstract the construction of error messages
with the {\sf \#errmessage} metafunction.

\fref{metatables_stats} describes how the metatable mechanism works for 
field updates over a non-table value or a missing key. Again, we 
make use of metafunctions that abstracts the inner 
workings of the metatable mechanism: \textsf{indexmetatable}, which looks for the metatable of its first argument and looks up the fallback
with the key passed as its second argument. 

The first two rules of \fref{metatables_stats} shows how this case
resolves differently depending on whether the handler is a function
or not (typically, in the second case the handler will be a table).
The last two rules show how the absence of handler has different
results depending on whether the original value is a table or not. 

\subsection{Semantics of programs and error handling}
\label{sec:semantics_programs}

\begin{figure}
\begin{tabular}{c}
\\
$\mathsf{\sigma\,\textbf{:}\,\theta\,\textbf{:}\,
	\Variable{Enp}[\![\;
	\Keyword{\$err}\;\Variable{v}
	]\!]\;
	\mapsto
	\sigma\,\textbf{:}\,\theta\,\textbf{:}\,
	\Keyword{\$err}\;\Variable{v}}$ \\
$\mathsf{\sigma\,\textbf{:}\,\theta\,\textbf{:}\,
	\Variable{E}[\![\; \LabelCons
	{\Variable{Enp}[\![\;\Keyword{\$err}\;\Variable{v}]\!]}{\Label{ProtMd}} \;]\!]\;
	\mapsto}~
  \mathsf{
	\sigma\,\textbf{:}\,\theta\,\textbf{:}\,
	\Variable{E}[\![\;
	\boldsymbol{<}\!\!\Keyword{false}, \Variable{v}\!\!\boldsymbol{>}
	]\!]}$ \\
$\mathsf{\sigma\,\textbf{:}\,\theta\,\textbf{:}\,
	\Variable{E}[\![\;
	\LabelCons
	{\Terminal{;}}
	{\Label{ProtMd}}
	\;]\!]\;
	\mapsto \sigma\,\textbf{:}\,\theta\,\textbf{:}\,
	\Variable{E}[\![\;
	\boldsymbol{<}\!\!\Keyword{true}\!\!\boldsymbol{>}
	]\!]}$ \\
$\mathsf{\sigma\,\textbf{:}\,\theta\,\textbf{:}\,
	\Variable{E}[\![\;
	\LabelCons{\boldsymbol{<}\!\!\Variable{v} \Terminal{,}\;...\!\!\boldsymbol{>}}}{\Label{ProtMd}}
\;]\!]\;
\mapsto
\sigma\,\textbf{:}\,\theta\,\textbf{:}\,
\Variable{E}[\![\;
\boldsymbol{<}\!\!\Keyword{true}, \Variable{v} \Terminal{,}\;...\!\!\boldsymbol{>}
]\!]$	\\[.4em]
\end{tabular}
\caption{Errors.}
\flabel{full_prog_err}
\end{figure}

The definition of the $\mapsto$ reduction relation that describes
the full semantics of Lua is essentially a straightforward
extension of the simpler relation given in Figure~10. The domain
now includes $\theta$, and maps the relations that
were described in previous sections. Each of these relations is extended
with the $\theta$ and $\sigma$ stores as needed. We omit these definitions for brevity.

More interestingly, in order to model the semantics of Lua's exception handling,
we must extend this relation. Lua's exception handling consist of two built-in
functions: {\sf error}, which throws an error (any Lua value, usually a {\sf
  string}), and {\sf pcall}, which executes a function in \emph{protected mode}. As any
other built-in function, this behavior can be override by a developer.

Normally an error aborts the program, but if it is thrown in the context of a
{\sf pcall}, it is caught. In that case, {\sf pcall} returns \false\ and the
error, otherwise, it returns {\sf true} and the values returned by the function
called.

\fref{full_prog_err} describes the part of the $\mapsto$ relation that models
error propagation and handling. For it, two new run-time constructs are added:
\Keyword{\$err} to denote an error, and
$\LabelCons{\Variable{s}}{\Label{ProtMD}}$ to denote code that must be executed
in protected mode. In the first rule, the evaluation context \Nonterminal{Enp} is identical to
\Nonterminal{E}, except that there are no occurrences of $\LabelCons{\Nonterminal{E}}{\Label{ProtMD}}$. The rule essentially aborts the whole program if there is no protected context around the
error. The second rule aborts up to the first  occurrence of a protected mode label,  if there is one. The other three
rules transition out of protected mode whether an error occurred or not.

The astute reader might wonder why we are modeling error handling here, and not as its own relation as we did with all the other parts of
the semantics. The reason is merely technical: the rule that aborts the whole program, if isolated in its own relation, would
break the {\em unique decomposition} property of evaluation contexts,
in which there is a single way for decomposing a term into an evaluation
context and the contents of its hole. We could have put just this rule explicitly in
$\mapsto$ while having the others in an hypothetical $\rightarrow^{\textsf{error}}$ relation, but decided to keep
all aspects of a feature together.

\section{Mechanization}
\label{sec:mechanization}

\begin{figure}
\begin{tabular}{|c|c|c|} \hline
\textbf{File}    & \textbf{Features tested}& \textbf{Coverage}\\ \hline
calls.lua        & functions and calls     & 77.83\%\\ \hline
closure.lua      & closures                & 48.5\%\\ \hline
constructs.lua   & syntax and              & 63.18\%\\ 
                 & short-circuit opts.     &        \\ \hline
events.lua       & metatables              & 90.4\%\\ \hline
locals.lua       & local variables         & 62.3\%\\ 
                 & and environments        &      \\ \hline
math.lua         & numbers and             & 82.2\%\\ 
                 & math lib                &       \\ \hline
nextvar.lua      & tables, next, and for   & 53.24\%\\ \hline
sort.lua         & (parts of) table        & 24.1\%\\ 
                 & library                 & \\ \hline
vararg.lua       & vararg                  & 100\%  \\ \hline
\end{tabular}
\caption{Lua 5.2's test suite coverage.}
\flabel{test_suite}
\end{figure}

The formalization of the semantics was carried in parallel with its
mechanization in PLT Redex~\cite{plt}.  This tool helped us recognize
problems in our first attempts at formalizing Lua, and allowed us to experiment
with new ideas before adding them to the formalization. It also allowed us to
execute part of test suite of the reference interpreter of the language\footnote{Available at https://www.lua.org/tests/.},
providing evidence that our semantics is in compliance with it.

Naturally, we could not use the whole test suite, for the following reasons:
\begin{compactitem}
\item Language features not covered by our formalization: coroutines, the \Keyword{goto} statement, and garbage collection;
\item Standard library functions not covered in our formalization: file
handling and string pattern matching. The Lua standard library is implemented
in C, so the parts we cover were ported by hand to the subset
of Lua covered by the formalization;
\item The \Keyword{debug} library, as it is heavily tied with the
implementation details of the interpreter;
\item Several other tests that test implementation details of the
interpreter, and not the language. According to the Lua authors, the
goal of the test suite is to test their reference {\em implementation} of Lua,
and not to serve as a conformance test for alternative implementations\footnote{
https://www.lua.org/wshop15/Ierusalimschy.pdf}.
\end{compactitem}

In practice, from the 25 \textsf{.lua} files present in the test suite,
which actually test some feature of the language, we are
able to port and run 9 against our PLT Redex mechanization. \fref{test_suite}
shows the percentage of LOCs actually tested from each of these remaining files, totaling
1256 LOCs successfully tested.  
It is important to remark that every file and line not tested is for the reasons
explained above, and every line (in the 9 files tested) that fall within the scope
of this work successfully passes the test. We take this as strong evidence that
the mechanization of our formal semantics behaves exactly the same as the
reference Lua interpreter.

Unfortunately, we do not have the space to discuss the code, which we plan to do
in an extended version of this article. For the moment, we refer the reader to
the documentation accompanying the code attached as supplementary material.

\paragraph*{Dynamic loading of code}
By implementing our parser directly in Racket (the language upon which
PLT Redex is based), we mechanized easily the \textsf{load}
service: Lua's compiler available at runtime.

There are several details to mention related to the solution implemented,
but for reasons of space we point out the most prominent:
\begin{itemize}
\item It covers the two modes: when the program to be compiled is passed as a
  string, or when it is a function from which the service obtains the program's
  string.
    
    \item It can handle the compilation of code on a modified
    global environment.
    
    \item For completion, we emulate the case of the compilation of
    \textit{binary chunks} (that is, a pre-compiled version of
    the code). This feature is implemented in conjunction with
    the service \textsf{string.dump}, which returns
    a string containing a binary representation
    of a function, given as a parameter.
\end{itemize}

\section{Related work}
\label{sec:relatedwork}

We mentioned already that the present development is inspired by the work done
in \cite{guh, get, python}, although not without its differences. For a start,
we tackle a smaller language, and use this fact to model its semantics
without shrinking much the core language. As a result,
we do not need to consider an external parser or interpreter: everything is
coded in Racket, and we can call the parser as a service of the library
(\sref{mechanization}). Another important difference is that we keep a certain
distance between formalization and mechanization: this is why our model of
memory is kept abstract, and why we put special emphasis in distinguishing the
run-time constructs from the constructs of the source language.

Specific to Lua's semantics, \cite{flua} presents the operational semantics of
Featherweight Lua, a minimal core of Lua. It considers a subset of features from
the ones presented here, and as
such cannot be tested against realistic Lua code. The mechanization is
implemented in Haskell, which is not as flexible as PLT Redex to extend and
change the model to cover a more realistic subset of the language.

There are several other important works related with real programming languages'
semantics.  K-Java~\cite{kjava} features a complete formal semantics for Java,
which is also mechanized using the tool $\mathbb{K}$. JavaScript, in turn, has
several formal semantics: besides the aforementioned \cite{guh, get},
\textsf{JSCert}~\cite{tjs} is a formalization of the language in the Coq proof
assistant, together with an interpreter extracted from the
formalization. \cite{opjs} also introduces a small-step operational semantics
for JavaScript, including proofs of several properties of the model.

\section{Conclusion}
\label{sec:conclusion}

We give a small-step operational semantics for a large subset of the Lua
programming language, as specified both by its informal reference manual and by
its reference implementation. We did so with little or no syntactical
differences between the actual syntax of Lua and the terms on which we define
the semantics, therefore having a lightweight elaboration or desugaring step.

The semantics tackles complex Lua features such as its metatable mechanism, dynamic execution of source code, error handling, and a several standard library 
functions that have no Lua implementation. It is defined in a modular way, and could be extended to tackle absent features such as coroutines
and the \Keyword{goto} statement without modifying what is already
specified.

We also provide a mechanization of the formal semantics in PLT Redex.
A large part of the test suite for the reference interpreter
has been successfully tested against this mechanization,
providing evidence that we are successfully modeling the
behavior of the language. As a plus, since our language resembles closely Lua, the traces of execution keep the code almost as written by the developer.

The development of the formal semantics, its mechanization, and its test suite
make up a tool that both semanticists and Lua developers can use for
understanding and extending the features of the language.

There are several further avenues for development. Besides adding missing
features such as coroutines and \Keyword{goto} statements, and the new operators
and large integer types of version 5.3, the formal semantics and its
mechanization can provide a basis for specifying, implementing, and formally
proving correct static analyses for Lua programs.

\bibliographystyle{plain}
\bibliography{references}
\end{document}